\documentclass[]{elsarticle}

\usepackage{lineno,hyperref}
\usepackage{ amsmath,amssymb}
\usepackage{graphicx,subcaption}% Include figure files
\usepackage{dcolumn}% Align table columns on decimal point
\usepackage{xcolor}
\usepackage{bm}% bold math
\usepackage{algorithm}
\usepackage{algorithmic}
\usepackage{amsmath}
\usepackage{subcaption}
\usepackage{scalerel}
\usepackage{tikz}
\usepackage{cleveref}
\usepackage{hyperref}
\usepackage{url}

\newcommand*\diff{\mathop{}\!\mathrm{d}}
\newcommand{\G}[3]{\ensuremath{\frac{1}{\sqrt{2\pi} #3} \exp{\Big\{ -\frac{(#1-#2)^2}{2#3^2} \Big\}}}}
\newcommand{\Go}[2]{\ensuremath{\frac{1}{\sqrt{2\pi} #2} \exp{\Big\{ -\frac{#1^2}{2#2^2} \Big\}}}}
\newcommand{\sigmatm}{\sigma(t_{\scaleto{\mathcal{M}}{3pt}})}

\newcommand{\m}{{\scaleto{\mathcal{M}}{3pt}}}
\newcommand{\w}{{\scaleto{\mathcal{W}}{3pt}}}

\journal{Journal of \LaTeX\ Templates}

\begin{document}

\begin{frontmatter}

\title{Limiting distribution of periodic position measurements of a quantum harmonic oscillator}

\author[]{Arnab Acharya\corref{auth1}}
%\ead{aa14ip031@iiserkol.ac.in}
\author[]{Debapriya Pal}

\author[]{Soumitro Banerjee\corref{auth3}}
\ead{soumitro@iiserkol.ac.in}

\author[]{Ananda Dasgupta\corref{auth4}}

\address{Department of Physical Sciences, Indian Institute of Science Education \& Research, Kolkata, Mohanpur Campus, Nadia 741246, WB, India}

\cortext[auth3]{Corresponding author}

\begin{abstract}
We consider a particle in harmonic oscillator potential, whose position is periodically measured with an instrument of finite precision. We show that the distribution of the measured positions tends to a limiting distribution when the number of measurements tends to infinity. We derive the expression for the limiting position distribution and validate it with numerical simulation.
\end{abstract}

%\begin{keyword}
%\texttt{elsarticle.cls}\sep \LaTeX\sep Elsevier %\sep template
%\MSC[****] **-**\sep  **-**
%\end{keyword}

\end{frontmatter}

%\linenumbers

\section{Introduction}

The operational approach to quantum mechanics \cite{busch1997operational,busch1996quantum,ozawa1984quantum} has, among other things, systematically expanded the notion of ideal projective measurements  to include imprecise and unsharp measurements. This has been fruitful for a number of practical \cite{konrad2012maintaining, choudhary2013implementation, katz2006coherent} as well as foundational problems \cite{ozawa1988measuring,busch1996insolubility}. 

 Unsharp measurements have been used to maintain coherence in the presence of noise \cite{konrad2012maintaining}. Choudhary \textit{et al}. \cite{choudhary2013implementation} have suggested their application in the measurement of qubit levels of a trapped ion. The evolution of a superconducting qubit subjected to unsharp measurement has been investigated \cite{katz2006coherent}. Schemes for reliable state estimation with sequential \cite{bassa2015process} and continuous-time unsharp measurements \cite{konrad2010monitoring} have been suggested. 

%cite all books related to measurements

%We have been able to derive closed form expression for a first approximation to the rigorous collapse scheme. Specifically, we assume: $\Psi \leftarrow^\text{measurement} \G{\mu}{}$

A special class of quantum measurements, called quantum non-demolition (QND) measurements have been widely used in monitoring a quantum oscillator \cite{braginsky1980quantum,thorne1978quantum}. 
This form of measurement can in principle leave the quantum state undisturbed. This could be  useful for extremely high precision measurements such as in certain schemes of gravitational wave detection. The statistical behavior of a quantum oscillator subjected to a sequence of QND measurements has been formally worked out by Matta and Pierro \cite{matta2015sequential}.   

Our aim in this work is to characterize the statistical distribution of a sequence of position measurements of a quantum oscillator. 

The interpretation of a wave function in the position basis is that
the absolute value of its square is the probability density of a
position measurement\cite{shankar2012principles}. Since a measurement in quantum mechanics changes
the state of the system, the notion of probability density is valid only in the
case of an ensemble of identically prepared states. In contrast,
in this work, we explore the consequences of periodic measurements on
the {\em same} quantum system.

The test system under study is the harmonic oscillator because of its
ubiquity and analytical ease. The system starts with a given
wavefunction. When an ideal position measurement is made, the wavefunction is supposed
to collapse to a delta function, whose position would be a random
number following the probability distribution given by the
wavefunction just before the collapse. Subsequently, the wavefunction
would evolve following the Schr\"odinger equation until the next
observation.  In order for the above scheme to work in numerical
simulation, we need the wavefunction to be smooth. So we consider the
state immediately after a collapse to be a narrow Gaussian function,
whose width represents the accuracy of measurement. In fact, it is supposed to collapse to a wavefunction which is a product of the wavefunction before measurement and the Gaussian function representing the measurement process. However, if the measurement process is represented by a Gaussian function that is narrow enough compared to the spread of the particle wavefunction, the post-collapse wavefunction can be aptly represented by the narrow Gaussian alone.

Using this schema, we show that the distribution of position
measurements tends to a limiting distribution in the limit of infinite
measurements. The limiting distribution is a Gaussian function centered
at zero, and with a standard deviation that depends on the frequency
of measurement, the width of the Gaussian function following a
collapse, and the characteristic parameters of the harmonic
oscillator. We obtain the limiting distribution in closed form and highlight some of its features. These results are validated using numerical simulation.

\begin{figure}
\centering
\includegraphics[width=0.45\linewidth]{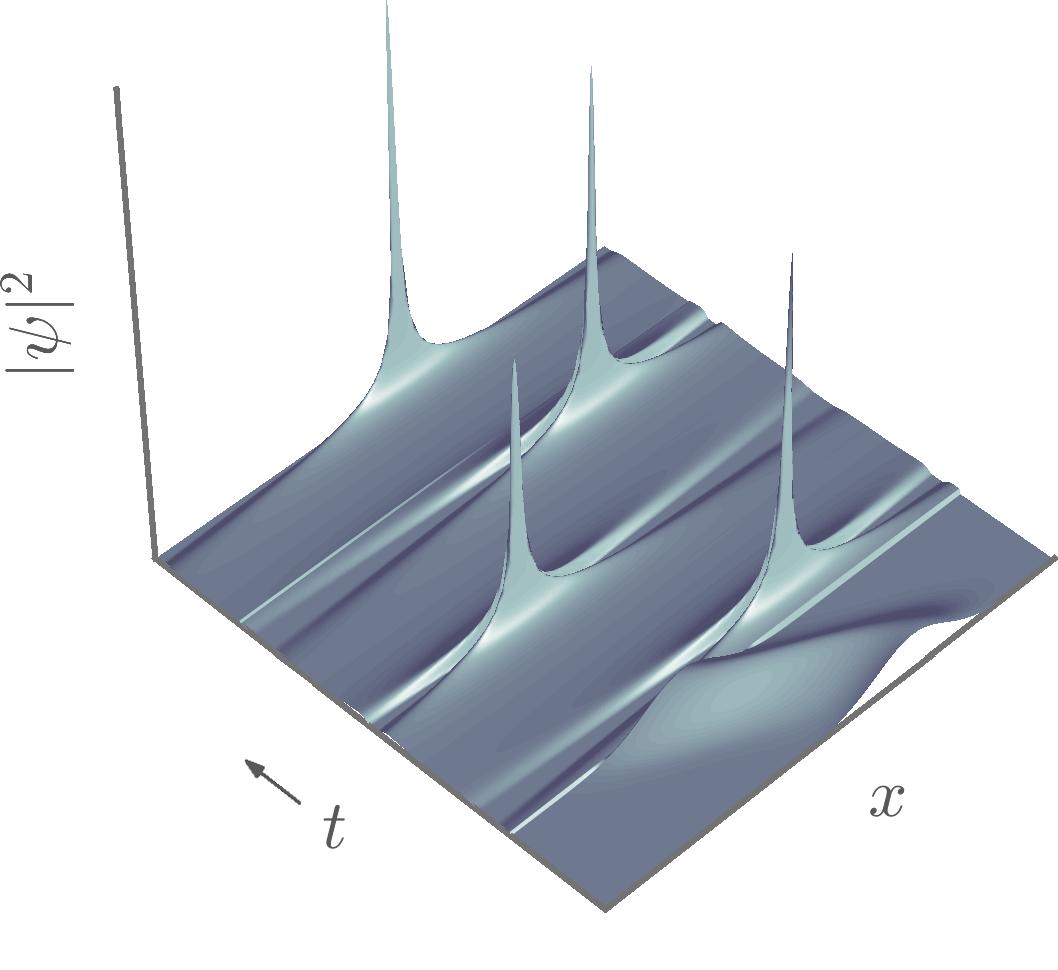}
\medskip
\caption{Evolution of the probability density function with periodic measurement}
\label{fig:SSoscl}
\end{figure}

\section{Problem Statement and Numerical Results}

A particle is placed in a harmonic oscillator potential 
\begin{equation}V = \frac{1}{2}m\omega^2x^2\end{equation} 
  The particle is initially in a state $\Psi(x,0)$ and we subject it
  to periodic measurements at intervals of time $t_\m$. 

  We assume that the measuring instrument has a finite
  precision so that the state of system collapses to a narrow Gaussian
  function centered at $x_\m$ with standard deviation $\sigma_\m$ after each measurement
  (this allows the wavefunction to be differentiable):
\begin{equation}\label{eq:scheme}
\Psi(x,t_\m)\xrightarrow{\text{measurement}} \mathcal{G}(x-x_\m,\sigma_\m).
\end{equation}
The case of
  ideal measurements is easily derivable by letting $\sigma_\m$ go to
  zero. 
  
  \begin{figure}[t]
\begin{subfigure}[b]{0.45\linewidth}
  \includegraphics[width=\linewidth]{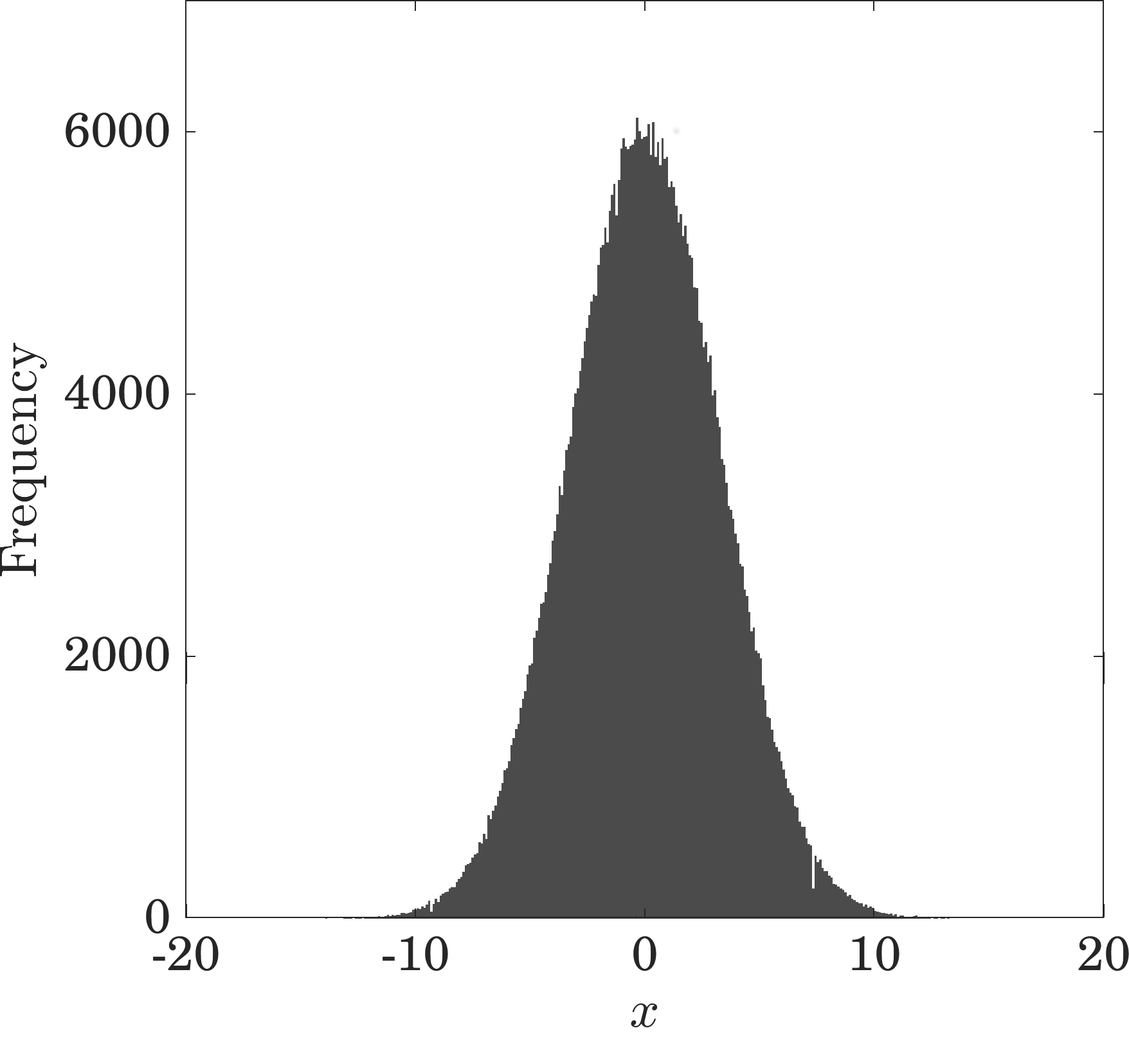}
  \caption{}
  \end{subfigure}
  \hspace{.5mm}
  \medskip
  \begin{subfigure}[b]{0.45\linewidth}
  \includegraphics[width=\linewidth]{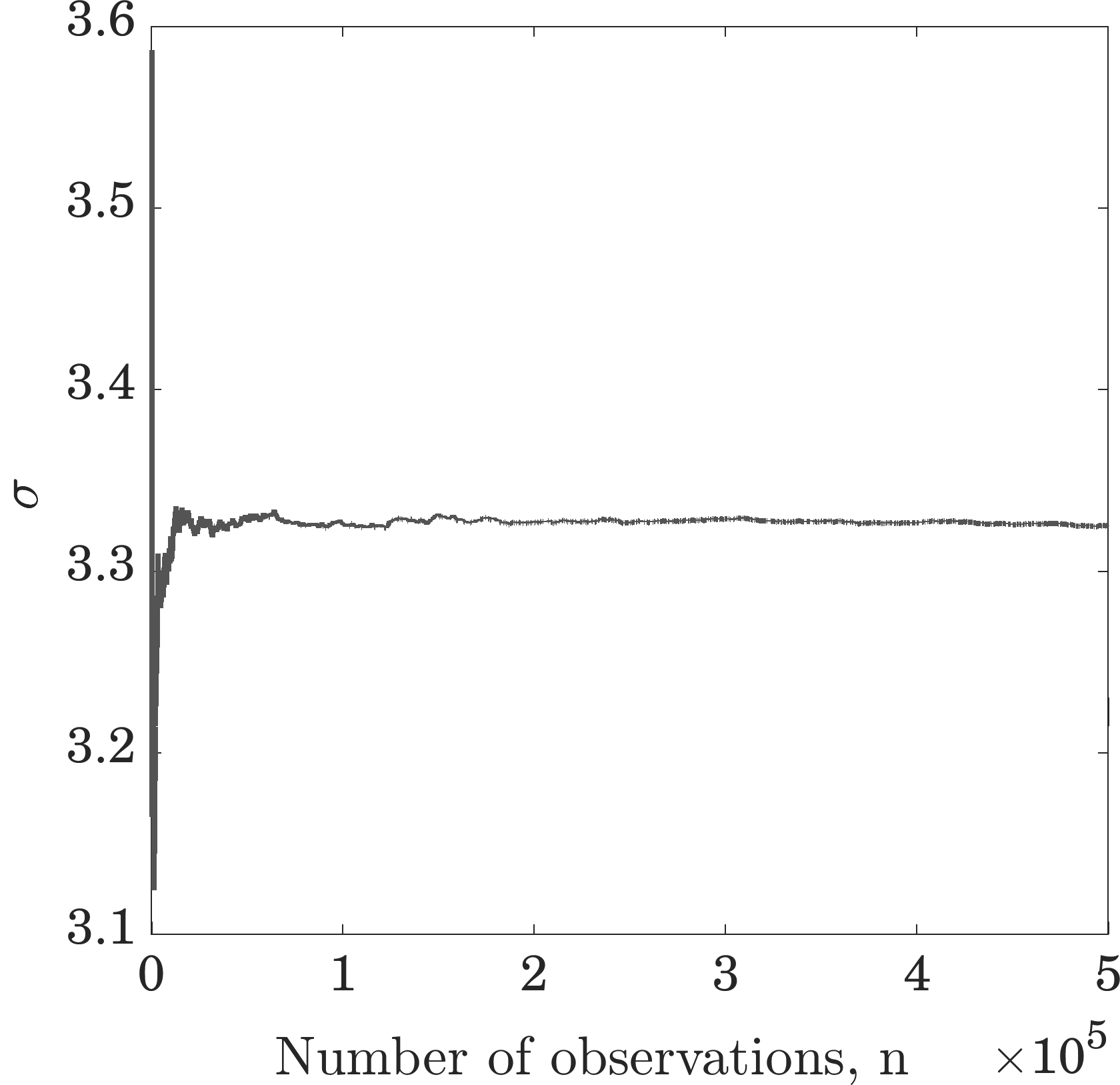}
  \caption{}
  \end{subfigure}
  \medskip
\caption{(a) Histogram of $5\times10^5$ measurements for $m = 1,\; \omega = 0.707,\; \sigma_\m = 0.5,\; t_\m = \frac{T}{5}$, (b) Convergence of the standard deviation, which tends to $\sigma_\infty$}
\label{fig:histogram}
\end{figure}

The system was simulated by evolving the wavefunction for $t_\m$ seconds and drawing a random sample $x_\m$ every $t_\m$ seconds from the probability density $|\Psi(x)|^2$ just before measurement. Just after the measurement it was replaced by a narrow Gaussian of standard deviation $\sigma_\m$ and centered at $x_\m$. The state was then allowed to evolve until the next measurement following the Schr\"odinger equation (Fig.~\ref{fig:SSoscl}). The process was repeated to obtain the limiting distribution of samples, which we plot in Fig.~\ref{fig:histogram}(a). Numerical simulations for various values of $t_\m$ and $\sigma_\m$ revealed that the limiting distribution is always Gaussian. The standard deviation of the samples was found to rapidly converge to a constant value as the number of samples was increased (Fig.~\ref{fig:histogram}(b)). This observation of convergence to a Gaussian distribution motivated our analytical approach, which is presented  in section \ref{sec:3}.

\begin{figure}
\centering
  \includegraphics[width=0.45\linewidth]{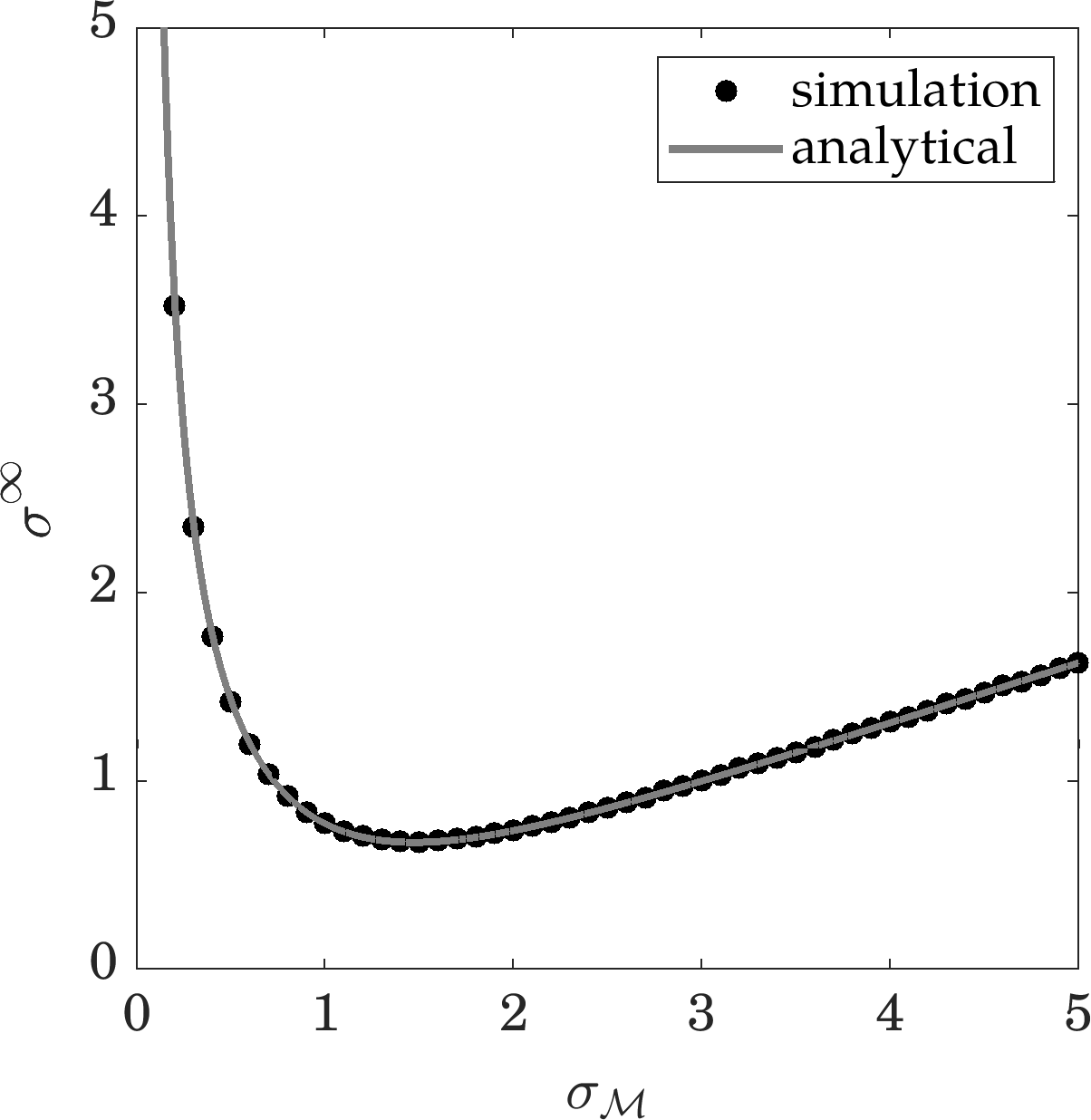}
  \includegraphics[width=0.45\linewidth]{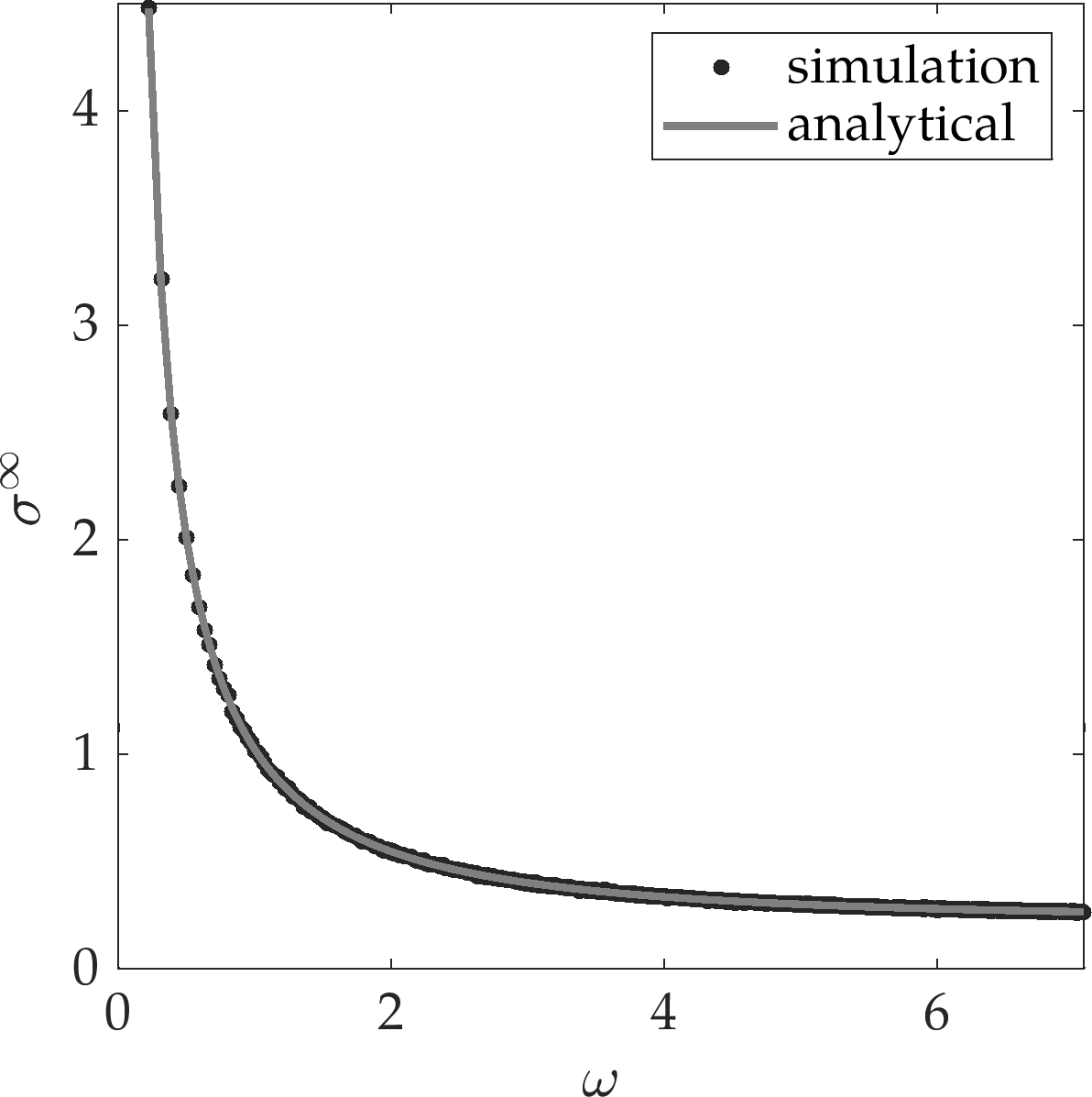}\\(a)\hspace{2in}(b)
  \medskip
  \caption{Limiting standard deviation versus (a) the standard deviation of collapsed wavefunction, and (b) the natural frequency of the harmonic oscillator.}
\label{fig:sigma_sigma}
\end{figure}

In order to check the dependence of the limiting standard deviation ($\sigma_\infty$) on the accuracy of the measuring device, we obtained the results for different values of $\sigma_\m$. The results are presented in Fig.~\ref{fig:sigma_sigma}(a). 
Similarly, we explore the dependence of the limiting standard deviation on the natural frequency of the harmonic oscillator, which we plot in Fig.~\ref{fig:sigma_sigma}(b). We also found that the limit distribution is independent of the initial wavefunction. 

The assumption of perfectly periodic measurements cannot be realized in practice. Hence, we numerically investigated the effect of a Gaussian noise added to the time period of measurement. We report that the additional noise has negligible effect on the limiting standard deviation. This implies that the analytical results presented below, which are derived based on the assumption of periodic measurements, also hold for approximately periodic measurements.

Note that our measurement scheme can be described in the language of positive operator-valued measures (POVMs) (see \ref{appendix:c}). Hence, the scheme is, in principle, realizable.

\section{Derivation of Limit Distribution}\label{sec:3}
We now obtain the expression for the limiting distribution and its dependence on various parameters. 

The evolution of a Gaussian wave packet in a harmonic potential is a well known result ~\cite{tsuru1991wave,saxon2013elementary,brandt2013quantum}. Let the initial wave packet be
\begin{equation}
\Psi(x,0) = \frac{1}{\sqrt[4]{2\pi} \sqrt{\sigma_{x0}} }\exp{\Big\{ -\frac{(x-x_0)^2}{4\sigma_{x0}^2} \Big\}}
\label{eq:gaussian}
\end{equation}
where $x_0$ is the initial center of the Gaussian wave packet,  $\sigma_{x0}$ is the initial width of wave packet. Then the probability density at time $t$ is given by
\begin{equation}
|\Psi(x,t)|^2 = \frac{1}{\sqrt{2\pi} \sigma(t)} \exp{\Big\{-\frac{(x-x_0\cos\omega t)^2}{2\sigma^2(t)}\Big\}}
\label{eq:timeGaussWP}
\end{equation}
where
\begin{equation}
\sigma(t) = \frac{\sigma_{gs}^2}{2\sqrt{2}\;\sigma_{x0}}\sqrt{4\left(\frac{\sigma_{x0}}{\sigma_{gs}}\right)^4+1+\left(4\left(\frac{\sigma_{x0}}{\sigma_{gs}}\right)^4-1\right)\cos{2\omega t}}
\label{eq:1}
\end{equation}
and $\sigma_{gs} = \sqrt{\hbar/m\omega}$ is the width of the ground-state eigenfunction.

%remove following 4 figures
%\begin{figure*}
 % \centering
  %\includegraphics[width=0.35\textwidth]{GS.png}%{\footnotesize (a)} 
  %\hspace{5mm}
  %\includegraphics[width=0.35\textwidth]{GS10.png}%{\footnotesize (b)}
  %\medskip
  %\\ (a) \hspace{1.7in} (b)  \\
  %\includegraphics[width=0.35\textwidth]{SS0.png}%{\footnotesize (c)} 
  %\hspace{5mm}
  %\includegraphics[width=0.35\textwidth]{SS10.png}%{\footnotesize (d)}
%\medskip
 % \\ (c) \hspace{1.7in} (d) 
  %\medskip
%\caption{(a) Stationary coherent wavepacket, (b) Oscillating coherent wavepacket, (c) Width oscillations of a squeezed wavepacket, and (d) General Squeezed wavepacket.}
%\label{fig:GS}
%\end{figure*}

For the sake of succinctness, we shall refer to a Gaussian in $x$ centered at $\mu$ with standard deviation $\sigma$ as $\mathcal{G}\left(x-\mu,\sigma\right)$, whereby the time evolution of a Gaussian wavepacket can be expressed as
\begin{equation}
\mathcal{G}\left(x-x_0, \sigma_{x0}\right)\xrightarrow{\;\;\;\; t \;\;\;\;}\mathcal{G}\left(x-x_0\cos{\omega t}, \sigma(t)\right)
\end{equation}

%Fig.~\ref{fig:GS} shows the distinct types of Gaussian wavepacket dynamics in the harmonic oscillator potential. 

At $t=0$ we start with a Gaussian wavepacket (\ref{eq:gaussian}) centered at $x=0$, and width $\sigma_{x0}$. We repeatedly measure the position of the particle after fixed time intervals of $t_\m$. At each measurement a random value of the position is chosen following the distribution of $|\Psi|^2$ at that time instant.

A measurement collapses the wavefunction. The imprecise instrument is assumed to collapse the wavefunction into a narrow Gaussian wavepacket
\begin{equation*}
\Psi_i(x,0) = \frac{1}{\sqrt[4]{2\pi} \sqrt{\sigma_\m}} \exp{\Big\{ -\frac{{(x-x_{\m i})}^2}{4{\sigma^2_\m}} \Big\}}
\end{equation*}
where $x_{\m i}$ is the outcome of the
$i^{\mathrm{th}}$ measurement. The next measurement happens after a
time $t_{\m}$. The probability density for the wavefunction just
before the next measurement can be calculated using equation
\eqref{eq:timeGaussWP}
\begin{align*}
|\Psi_i(x,t_\m)|^2 &= \frac{1}{\sqrt{2\pi} \sigma(t_\m)} \exp{\Big\{-\frac{(x-x_{\m i}\cos\omega t_\m)^2}{2\sigma^2(t_\m)}\Big\}}\\ &= \mathcal{G}\left(x-x_{\m i} \cos\omega t_\m,\sigma(t_\m)\right)
\end{align*} 

\noindent For the first measurement the distribution is  
\begin{align*}
\mathcal{D}_1(x) &= |\Psi(x,t_{\m})|^2\\ &=  \frac{1}{\sqrt{2\pi} \sigma_0(t_{\m})} \exp{\Big\{-\frac{x^2}{2\sigma_0^2(t_{\m})}\Big\}}\\ &= \mathcal{G}\left(x,\sigma_0(t_{\m})\right)
\end{align*}

We denote the standard deviation of this distribution as $\sigma_0(t_\m)$ to distinguish it from all subsequent standard deviations. The densities before all subsequent measurements have the same width as they all start from a collapsed state whose standard deviation is identical in all cases. The expected distribution for the second measurement is

\begin{eqnarray*}
  \mathcal{D}_2(x) &=& \int_{-\infty}^{\infty}\mathcal{D}_1(x_{\m 1})\;|\Psi_1(x,t_\m)|^2 \diff x_{\m 1}\\  
& =& \mathcal{G}\left(x,\sigma(t_\m) \sqrt{1 + \left(\frac{\sigma_0(t_\m)}{\sigma(t_\m)}\right)^2\cos^2{\omega t_\m}}\right)
\end{eqnarray*}

\noindent The derivation of this result can be found in the Appendix \ref{appendix:a}.
 Similarly, for the third measurement, the density is

\begin{align*}
\mathcal{D}_3(x) &= \int_{-\infty}^{\infty}\mathcal{D}_2(x_{\m 2})\;|\Psi_2(x,t_\m)|^2 \diff x_{\m 2}\\
& = \mathcal{G}\left(x,\sigma(t_\m)\sqrt{1 + \cos^2{\omega t_\m}+ \left(\frac{\sigma_0(t_\m)}{\sigma(t_\m)}\right)^2\cos^4{\omega t_\m}}\right)
\end{align*}

\noindent And for the $n^{\text{th}}$ measurement the density is

\begin{align*}
\mathcal{D}_n(x) &= \mathcal{G}\left(x,\sigma(t_\m)\sqrt{1 + cos^2{\omega t_\m}+ \dots+ \left(\frac{\sigma_0(t_\m)}{\sigma(t_\m)}\right)^2 \cos^{2(n-1)}{\omega t_\m}}\right)
\end{align*}

\noindent The geometric series converges if $\cos^2{\omega t_\m} < 1$

\begin{align}
& &\mathcal{D}_\infty(x) &= \mathcal{G}\left(x,\sigma(t_\m)\sqrt{1 + cos^2{\omega t_\m}+ \dots}\right)\nonumber\\\nonumber
&&&= \mathcal{G}\left(x,\sigma(t_\m)\sqrt{\frac{1}{1 - cos^2{\omega t_\m}}}\right)\\&&&= \mathcal{G}\left(x,\frac{\sigma(t_\m)}{\sin{\omega t_\m}}\right)\\
&\text{or,} &\sigma_\infty &= \left|\frac{\sigma(t_\m)}{\sin{\omega t_\m}}\right|
\end{align}

The distribution of position measurements is then the distribution of samples, one taken from each $\mathcal{D}_i$. This is the mean of the densities $\mathcal{D}_i$. So we have

\begin{equation}
\sigma_i^2 = \sigma(t_\m)^2 + \sigma(t_\m)^2 cos^2{\omega t_\m}+ \dots+ \sigma_0(t_\m)^2 \cos^{2(i-1)}{\omega t_\m}
\label{eq:var}
\end{equation}

The mean of all these densities is again a Gaussian with mean at zero and variance $s_\infty^2$ given by the mean of the individual variances given by equation (\ref{eq:var}). 

$$s_\infty^2 =\lim_{n\rightarrow\infty}\frac{1}{n}\sum_{i = 1}^\infty \sigma_i^2\\ = \left\{\frac{\sigma(t_\m)}{\sin{\omega t_\m}}\right\}^2 $$

\noindent The calculation of this result is given in Appendix \ref{appendix:b}.

\begin{eqnarray}
\therefore\; s_\infty&=&\left|\frac{\sigma(t_\m)}{\sin{\omega t_\m}}\right| = \sigma_\infty
\label{eq:s_inf}
\end{eqnarray}

 We find that $s_\infty$ is the same as $\sigma_\infty$. This means that the distribution of measurement outcomes $x_{\m 1},x_{\m 2},\dots,x_{\m n}$ itself converges to $\mathcal{D}_\infty$ as $n\rightarrow\infty$.

Substituting equation (\ref{eq:1}) in equation (\ref{eq:s_inf}) we get
\begin{equation*}
\sigma_\infty = \left|\frac{\sigma_0^2\sqrt{4\big(\frac{\sigma_{x0}}{\sigma_{0}}\big)^4+1+\big(4\big(\frac{\sigma_{\m}}{\sigma_{0}}\big)^4-1\big)\cos{2\omega t_{\m}}} }{2\sqrt{2}\;\sigma_{\m}\sin{\omega t_{\m}}}\right|
\end{equation*}
 
%change notation in fig 3

In Fig.~\ref{fig:sigma_sigma} the analytical result given by equation~(\ref{eq:s_inf}) is plotted with continuous lines while the numerical results are plotted with dots. In both these cases the theoretical results and simulation show good agreement. 

\begin{figure}
\centering
\includegraphics[width=0.45\linewidth]{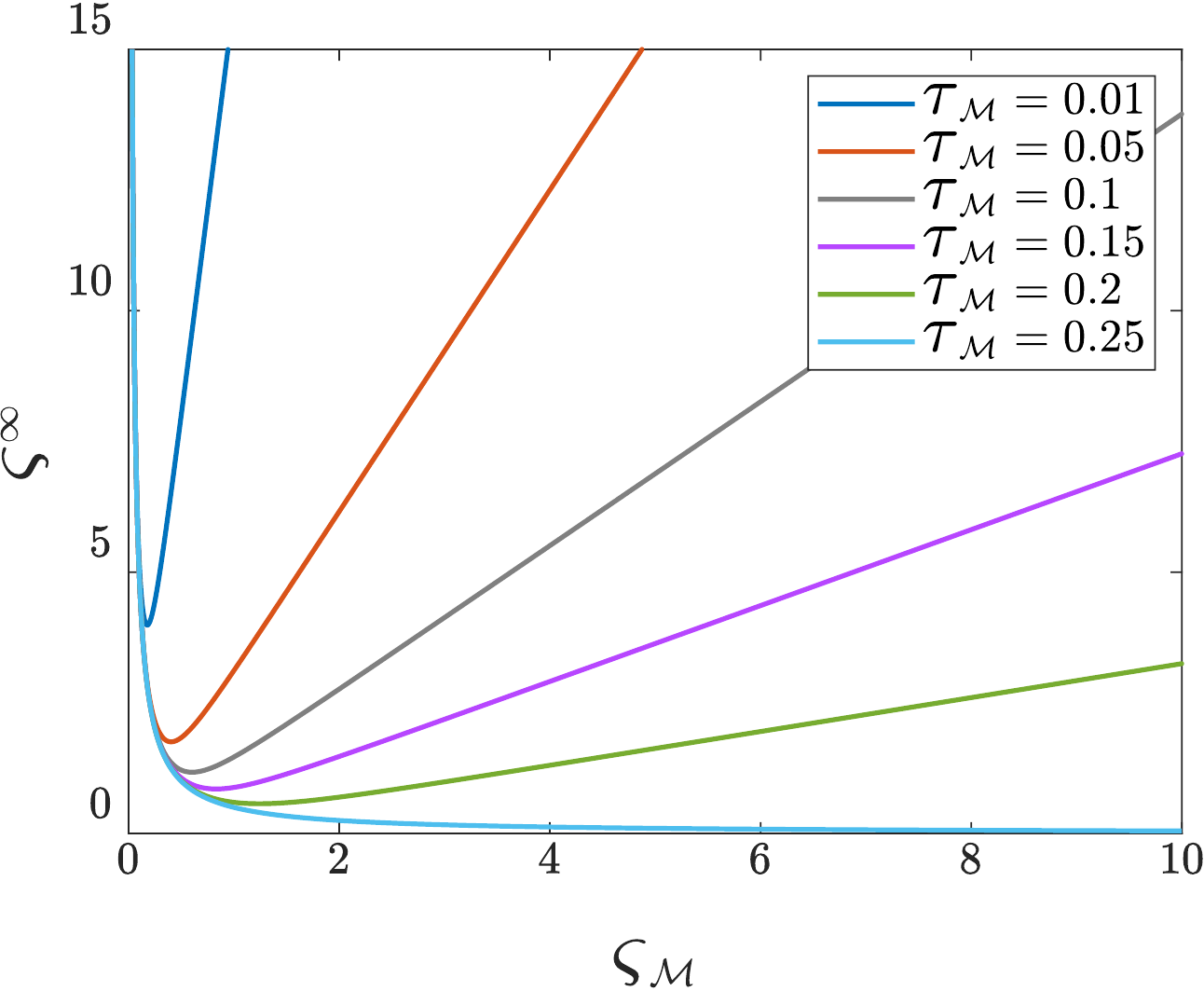}\hspace{0.3in}
\includegraphics[width=0.45\linewidth]{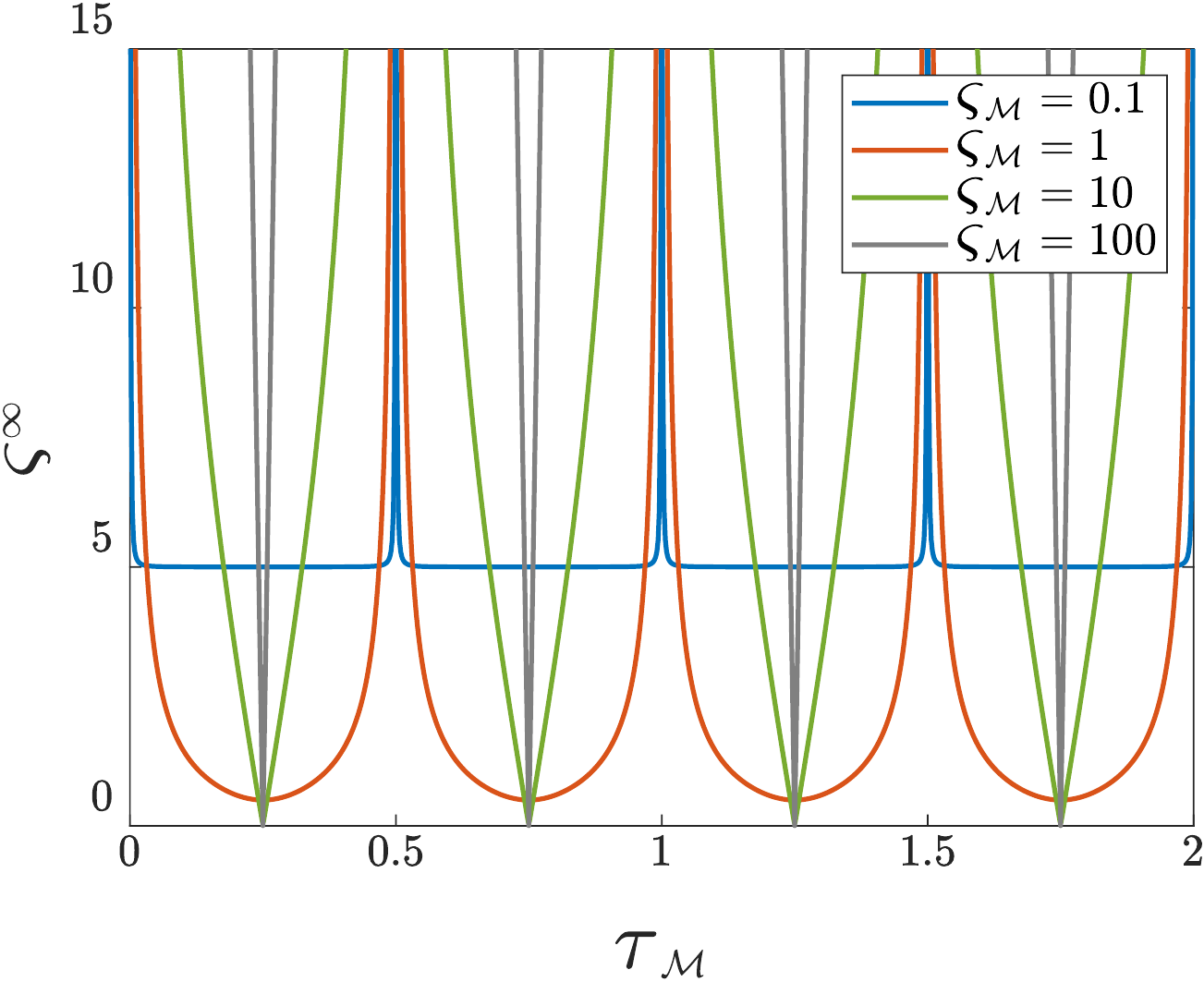}\\\hspace{0.4in}(a)\hspace{3.15in}(b)
  \medskip
\caption{Plots of $\varsigma^{\scaleto{\infty}{2pt}}$ vs (a)  $\varsigma_\m$ and (b)  $\tau_\m$. Color online.}

\label{fig:var}
\end{figure}

\section{Analysis}

We found the limit distribution to be a Gaussian centered at zero with standard deviation given by equation (\ref{eq:s_inf}) which can be simplified to
\begin{equation*}
\sigma_\infty = \sqrt{\sigma_{\m}^2 \cot^2\omega t_{\m}+\frac{\sigma_0^4}{4\sigma_{\m}^2}}
\end{equation*}

\noindent We obtain a non-dimensional form by dividing throughout by $\sigma_0$,

\begin{align*}
&&\frac{\sigma_\infty}{\sigma_0} &= \sqrt{\left(\frac{\sigma_{\m}}{\sigma_0}\right)^2 \cot^2 2\pi\frac{ t_{\m}}{T}+\frac{1}{4}\left(\frac{\sigma_0}{\sigma_{\m}}\right)^2}\\
&\text{or,} &\varsigma^{\scaleto{\infty}{2pt}} &= \sqrt{\varsigma_{\m}^2 \cot^2 2\pi \tau_{\m} + \frac{1}{4\varsigma^2_{\m}}}
\end{align*}
where $\varsigma^{\scaleto{\infty}{2pt}} = \frac{\sigma^{\scaleto{\infty}{2pt}}}{\sigma_0}$, $\varsigma_\m = \frac{\sigma_\m}{\sigma_0}$ and $\tau_\m = \frac{t_m}{T}$. These substitutions are advantageous because they are dimensionless quantities independent of the length and time-scales of any particular harmonic oscillator. 

In Fig.~\ref{fig:var}(a) we see how $\varsigma^{\scaleto{\infty}{2pt}}$ changes when we vary $\varsigma_\m$, for particular values of $\tau_\m$. For $\varsigma_\m \rightarrow 0$, $\varsigma^{\scaleto{\infty}{2pt}}$  grows hyperbolically ($\sim\frac{1}{2\varsigma_\m}$). It has a minimum value at $\varsigma_\m = \sqrt{\frac{\tan 2\pi\tau_{\scaleto{\m}{2.5pt}}}{2}}$. And as $\varsigma_\m \rightarrow \infty$, it grows linearly in $\varsigma_\m$ with slope $\left|\cot 2\pi\tau_\m\right|$. For $\tau_\m \in \left[0,\frac{1}{4}\right]$ the slope of the linear asymptote varies from $\infty$ to $0$. After $\tau_\m = \frac{1}{4}$, the process reverses itself till $\tau_\m = \frac{1}{2}$, after which the pattern repeats periodically. 

In Fig.~\ref{fig:var}(b) we see how $\varsigma^{\scaleto{\infty}{2pt}}$ changes with $\tau_\m$. The plots are periodic in $\tau_\m$ with period $\frac{1}{2}$. The curves have minima at $\frac{n}{2} - \frac{1}{4}, n \in \mathbb{N}$. As $\varsigma_\m$ increases $\varsigma^{\scaleto{\infty}{2pt}}$ gets steeper and the minima tend to zero.
 
\begin{figure}
\centering
\includegraphics[width=0.5\linewidth]{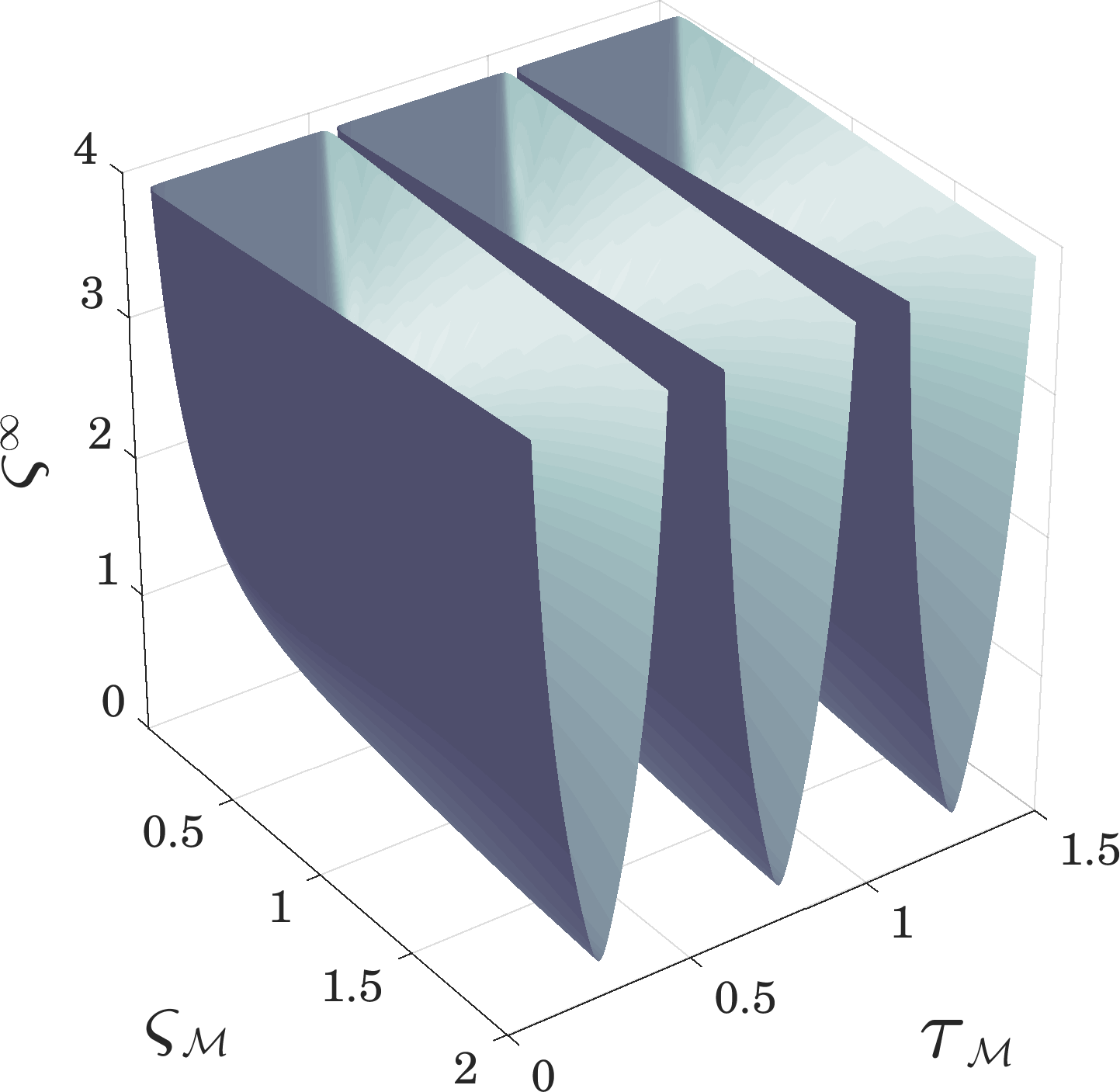}
\caption{Plot of $\varsigma^{\scaleto{\infty}{2pt}}$ vs  $\varsigma_\m$ and $\tau_\m$}
\label{fig:var_3d}
\end{figure}

In Fig.~\ref{fig:var_3d} the dependence of $\varsigma^{\scaleto{\infty}{2pt}}$ on both $\tau_\m$ and $\varsigma_\m$ have been consolidated into a single surface plot.

\section{Conclusion}

We have investigated the statistical distribution of periodic measurements on a single quantum system (in this case a quantum harmonic oscillator). We find that the measurement outcomes follow a Gaussian distribution with mean zero.  

An analytical expression for the standard deviation of the limiting distribution was derived and was validated with numerical simulation. 
The standard deviation of this distribution was found to depend on the accuracy and frequency of measurements, and the natural frequency of the harmonic oscillator. This distribution was found to be independent of the initial wavefunction.

We have shown that there is an optimal accuracy of measurement that minimizes the standard deviation of the limit distribution.  We also found  that certain measurement intervals minimize the standard deviation of the limit distribution. The analytical results were numerically found to be robust to the presence of Gaussian noise in the time interval. These results may be useful for localizing a particle at the center of a well with the least uncertainty.

%\section*{Declaration of competing interest}The authors declare that they have no known competing financial interests or personal relationships that could have appeared to influence the work reported in this paper.

\section*{Acknowledgement}
AA acknowledges the financial support from Institute Fellowship of
IISER Kolkata. SB acknowledges the J C Bose National Fellowship
provided by SERB, Government of India, Grant No. SB/S2/JCB-023/2015.

\section*{References}
\bibliography{mybibfile}

\appendix
\section{A POVM reformulation of our measurement scheme}
\label{appendix:c}
We shall demonstrate that our scheme can be realized as a generalized measurement (POVM). In particular, we show that our scheme can be reformulated as a weak position measurement, which is a subclass of generalized measurements \cite{gross2015novelty}. 

The measurement scheme used in this work (\ref{eq:scheme}) is
$$\Psi(x,t_\m)\xrightarrow{\quad\text{measurement}\quad} \mathcal{G}(x-x_\m,\sigma_\m).$$
  which, we shall show, is a special case of the weak measurement protocol used in \cite{matta2015sequential,konrad2010monitoring}
$$\Psi(x)\xrightarrow{\text{measurement}} \frac{1}{\mathcal{N}} \exp\left\{\frac{-(x-x_\w)^2}{2\sigma_\w^2}\right\}~\Psi(x),$$
where $\sigma_\w$ is a measure of the  imprecision of the instrument and $x_\w$ denotes the position registered by the instrument. $\mathcal{N}$ ensures normalization of the post-measurement state. 

Equating the post-measurement states of both these schemes,
$$\mathcal{G}(x-x_\m,\sigma_\m) = \frac{1}{\mathcal{N}}~\exp\left\{\frac{-(x-x_\w)^2}{2\sigma_\w^2}\right\}~\Psi(x)$$
Since the states used in this work are Gaussian wave packets, $\Psi(x)$ can be written as $\mathcal{G}(x-\mu_\psi,\sigma_\psi)$, which gives

\begin{equation*}\label{eq:postm}
\mathcal{G}(x-x_\m,\sigma_\m) = \frac{1}{\mathcal{N}}~\mathcal{G}(x-x_\w,\sigma_\w)~\mathcal{G}(x-\mu_\psi,\sigma_\psi)
\end{equation*}

\noindent Using results about the product of two Gaussians \cite{bromiley2003products}, we write

$$\sigma_\m^2 = \frac{\sigma_\w^2\sigma_\psi^2}{\sigma_\w^2+\sigma_\psi^2} \text{\qquad  \qquad} x_\m=\frac{x_\w\sigma_\psi^2 + \mu_\psi\sigma_\w^2}{\sigma_\w^2+\sigma_\psi^2}$$

%$$\mathcal{N} = \frac{1}{\sqrt{2\pi(\sigma_\w^2+\sigma_\psi^2)}}~\exp\left\{\frac{(x_\w - \mu_\psi)^2}{2(\sigma_\w^2+\sigma_\psi^2)}\right\}$$

\noindent Rearranging terms,
$$\sigma_\w^2 = \frac{\sigma_\m^2\sigma_\psi^2}{\sigma_\psi^2-\sigma_\m^2} \text{\qquad  \qquad} x_\w=\frac{x_\m\sigma_\psi^2 - \mu_\psi\sigma_\w^2}{\sigma_\psi^2-\sigma_\m^2}$$

Hence, our measurement scheme is equivalent to a weak measurement with the above parameters.

\section{Derivation of $\mathcal{D}_2(x)$}
\label{appendix:a}
\begin{align*}
  \mathcal{D}_2(x) &= \int_{-\infty}^{\infty}\mathcal{D}_1(x_{\m 1})\;|\Psi_1(x,t_\m)|^2 \diff x_{\m 1}\\
    &= \int_{-\infty}^{\infty}\mathcal{G}\left(x_{\m 1},\sigma_0(t_\m)\right)\;\mathcal{G}\left(x-x_{\m 1}\cos\omega t_\m,\sigma(t_\m)\right)\diff x_{\m 1} \\
  & =  \int_{-\infty}^{\infty} \Go{(x_{\m 1})}{\sigma_0 (t_\m)}\\
&\qquad\G{x}{x_{\m 1}\cos\omega t_\m}{\sigmatm}\diff x_{\m 1}
\\ & =  \int_{-\infty}^{\infty} \Go{(x_{\m 1})}{\sigma_0(t_\m)}\sec\omega t_\m\\
  &\quad\G{x_{\m 1}}{x\sec\omega t_\m}{(\sigmatm\sec\omega t_\m)}\diff x_{\m 1}\\
& =\int_{-\infty}^{\infty}\sec\omega t_\m\mathcal{G}\left(x_{\m 1},\sigma_1(t_\m)\right)~\mathcal{G}\left(x_{\m 1} - x\sec\omega t_\m,\sigma(t_\m)\sec\omega t_\m\right)\diff x_{\m 1}\\
\intertext{Using the following result about the integral of the product of two Gaussians}
&\int_{-\infty}^{\infty}\mathcal{G}\left(x-\mu_1,\sigma_1\right)\;\mathcal{G}\left(x-\mu_2,\sigma_2\right)\diff x = \quad\mathcal{G}\left(\mu_1-\mu_2,\sqrt{\sigma_1^2 + \sigma_2^2}\right)
\intertext{we have}
\mathcal{D}_{2}(x) 
 & = \sec\omega t_\m\mathcal{G}\left(x\sec\omega t_\m,\sqrt{\sigma_0(t_\m)^2 + \sigma(t_\m)^2\sec^2{\omega t_\m}}\right)
\\ & = \frac{\sec\omega t_\m \exp{\Big\{\scaleto{-\frac{x^2\sec^2\omega t_\m}{2\left\{\sigma(t_\m)^2\sec^2{\omega t_\m} + \sigma_0(t_\m)\right\}^2}}{20pt}\Big\}}}{\sqrt{2\pi}\sqrt{\sigma(t_\m)^2\sec^2{\omega t_\m} + \sigma_0(t_\m)^2}}\\
 & = \frac{ \exp{\Big\{\scaleto{-\frac{x^2}{2\left\{\sigma(t_\m)^2 + \sigma_0(t_\m)\cos^2{\omega t_\m}\right\}^2}}{20pt}\Big\}}}{\sqrt{2\pi}\sqrt{\sigma(t_\m)^2 + \sigma_0(t_\m)^2\cos^2{\omega t_\m}}}\\
& = \mathcal{G}\left(x,\sigma(t_\m) \sqrt{1 + \left(\frac{\sigma_0(t_\m)}{\sigma(t_\m)}\right)^2\cos^2{\omega t_\m}}\right)\\
\end{align*}

%\boxalign[0.6\columnwidth]{\begin{align*}
%\mathcal{D}_2(x) = \mathcal{G}\left(x,\sigma(t_\m) \sqrt{1 + %\left(\frac{\sigma_0(t_\m)}{\sigma(t_\m)}\right)^2\cos^2{\omega t_\m}}\right)
%\end{align*}}

\section{Calculating $s_\infty$}
\label{appendix:b}
\begin{eqnarray*}
s_\infty^2 &=& \lim_{n\rightarrow\infty}\frac{1}{n}\sigma_0(t_\m)^2\\
&&+\frac{1}{n}\left\{\sigma(t_\m)^2 + \sigma_0(t_\m)^2 cos^2{\omega t_\m}\right\}\\
&&\qquad\vdots\\
&&+\frac{1}{n}\left\{\sigma(t_\m)^2 + cos^2{\omega t_\m}+ \dots+ \sigma_0(t_\m)^2\cos^{2(i-1)}{\omega t_\m}\right\}\\
&&\qquad\vdots\\
&=&\lim_{n\rightarrow\infty}\Bigg[\quad \frac{\sigma_0(t_\m)^2}{n}\left\{ 1+\cos^2\omega t_\m + \cos^4\omega t_\m +\dots \right\}\\
  &&\qquad+\quad\frac{\sigma(t_\m)^2}{n}[n + (n-1)\cos^2{\omega t_\m}+ \dots\\
    &&\qquad+ \quad(n-(i-1))\cos^{2(i-1)}{\omega t_\m}+\dots\quad\Bigg]\\ 
  &=&\lim_{n\rightarrow\infty}\frac{\sigma_0(t_\m)^2 \text{cosec}^2\omega t_\m}{n}
  + \frac{\sigma(t_\m)^2}{n}\sum_{k = 1}^{\infty}[n-(k-1)](\cos^{2}{\omega t_\m})^{(k-1)}
\end{eqnarray*}

If $\cos^2{\omega t_\m}<1$ the second term is a convergent arithmetico-geometric series and we have
\begin{eqnarray*}
s_\infty^2 &=&\lim_{n\rightarrow\infty}\left[\frac{\sigma_0(t_\m)^2 \text{cosec}^2\omega t_\m}{n}+\frac{\sigma(t_\m)^2}{n} \left\{\frac{n}{1-\cos^2{\omega t_\m}}-\frac{\cos^2{\omega t_\m}}{(1-\cos^2{\omega t_\m})^2}\right\}\right]\\
\therefore\; s_\infty^2&=&\left\{\frac{\sigma(t_\m)}{\sin{\omega t_\m}}\right\}^2
\end{eqnarray*}

\end{document}